\documentstyle[times,DINA4,12pt]{article}
\newcommand{\be}{\begin{equation}}
\newcommand{\ee}{\end{equation}}
\newcommand{\bea}{\begin{eqnarray}}
\newcommand{\eea}{\end{eqnarray}}
\newcommand{\rmi}{i}
\newcommand{\geff}{\mbox{$ g_{eff} (\sigma)$}}

\newcommand{\vx}{{\bf x}}
\newcommand{\vp}{{\bf p}}

\begin{document}
\begin{titlepage}

\title{\bf A transport theory of relativistic nucleon--nucleon collisions
with confinement
\footnote{Part of the dissertation of
	  Thomas Vetter}
\footnote{Supported by BMFT and GSI Darmstadt}}
\author{T.\ Vetter,
T.\ S.\ Bir\'{o}, and U.\ Mosel
\\Institut f\"ur Theoretische Physik
, Universit\"at Giessen\\
35392 Giessen, West Germany}
\date{}

\maketitle
\pagestyle{empty}

\begin{abstract}
A transport theory is developed on the quark level to describe
nucleon--nucleon collisions. We treat the strong interaction effectively by
the Friedberg-Lee model both in its original and in its
modified confining version.
First we study the stability of the static three-dimensional semiclassical
configuration, then we present results of
the time evolution
given by a Vlasov equation for the quarks coupled to a Klein-Gordon equation
for the mean field. We find at higher energies that the nucleons are almost transparent, whereas
at lower energies we observe a substantial interaction. At
very low energies we see a fusion of our bags, which is due to the
purely attractive nature
of the mean field and hence is an artifact of our model.
We test the confinement mechanism and find that at higher
energies the nucleons are restored shortly after the collision.
\end{abstract}
\end{titlepage}

\pagestyle{plain}
\section{Introduction} \label{Intro}
The success of kinetic theories in describing intermediate energy heavy
ion collisions (\cite{CaMo90} --
\cite{Boter}) has stimulated the development of transport theories
based on QCD to describe collisions at very high energies
(\cite{Elze1}-- \cite{Stan}). However the complexity
of the resulting equations has so far prevented their application in actual
calculations. On the other hand
for bombarding energies larger than $\approx 100
\, GeV$ QCD can be treated perturbatively. This is the motivation for many parton cascade 
models (Fritjof \cite{AnGuNi87}, Venus \cite{We89}, RQMD \cite{SoStGr89/2}, HIJING
\cite{WaGy91}, parton cascade \cite{GeMu92}).

We are interested in the energy range of a few GeV where nonperturbative
QCD effects are still important.
In a recent article \cite{KaVeBiMo93}
we presented the derivation of such a transport theory and showed the
results of a one-dimensional calculation. In this paper we extend the above
method to three spatial dimensions. Due to the sensitive dependence of soliton
properties on
the number of spatial dimensions we put some emphasis on the study of the
stability of the solitonic solution.
Zhang and Wilets
\cite{Wil} have also derived transport equations based on the
Nambu Jona--Lasinio model in order to estimate chiral symmetry effects in heavy ion
collisions; this  is conceptionally close to our work. However, besides the absence of 
confinement in their model, these authors do not actually perform dynamical simulations.
Also soliton-soliton collisions based on the Skyrme-model
\cite{VeWaWaWy87},\cite{AlKoSeSo87} and on a
Skyrme-like $\sigma$-model \cite{KuPiZa93} have been performed in the past.
However these calculations do not include any explicit quark degrees of
freedom.

We assume that the nonperturbative effects of QCD can be modeled by the
exchange of a scalar field, as it is successfully done in the well known
Friedberg-Lee Model (for a review see \cite{Wilbuch}) .
Since in this model the hadron surface is dynamically generated it allows for a time
dependent dynamical treatment.

The outline of the paper is as follows:
In section \ref{FBL} we survey the static Friedberg-Lee
model and focus especially on its absolute confining extension. In section
\ref{transport} we show the
equations of motion for the phase-space evolution of the quarks in this model.
In section \ref{static} we present the result of our static solution and
study the stability of the soliton. In section \ref{collisions} we present
results of three-dimensional simulations. Finally we conclude and
summarize in section \ref{summary}. The appendix contains some details
of the numerical implementation.

\section{The Friedberg--Lee Model} \label{FBL}

The nontopological soliton model of Friedberg and Lee \cite{FbLe77,FbLe78},
further developed by Goldflam and Wilets \cite{GoWi82}, has enjoyed a considerable
interest in recent years, because it gives a dynamical description of
nucleons as solitons, bound by a self-interacting mean field.
Solutions of this model have been extensively studied during the last decade
\cite{Wilbuch},\cite{WiBiLuHe85} -- \cite{BiBiWi87}. In the original version,
without perturbatively coupled vector fields, the
Lagrangian is given by
\be
{\cal L} = \bar{\Psi} \left( \rmi \gamma_\mu \partial^\mu - m_0 \right) \Psi -	\bar{\Psi}
g (\sigma) \Psi + \frac{1}{2} \left(\partial_\mu \sigma \right)^2 - U (\sigma) \, ,
\label{FBLmod}
\ee
with
\be
\label{eq:coupling}
g(\sigma) = g_0 \, \sigma \, ,
\ee
which describes quarks with a rest mass $m_0$ coupled to a scalar field $\sigma$,
which in turn is assumed to mimic long-range nonabelian effects of many
gluon exchanges. The self-interaction potential is supposed to be quartic in
the field,
\be
\label{eq:u}
U(\sigma) = \frac{a}{2!} \sigma^2 +\frac{b}{3!} \sigma^3 + \frac{c}{4!} \sigma^4 + B   \, ,
\ee
and the constants in (\ref{eq:u}) are adjusted such that $U(\sigma)$ has a
minimum at $\sigma=0$ and another, energetically lower one, at
a nonzero vacuum expectation value. Here the potential is assumed to vanish.
Thus the constant
$U(\sigma) = B$ gives the energy density associated with a cavity with
$\sigma =0$ in the  nonperturbative QCD vacuum characterized by the
nonvanishing field $\sigma_{vac}$.
Confinement in the Friedberg-Lee model is realized through the
coupling of the fermions to the $\sigma$-field, providing an effective mass
for the quarks, which is larger in the vacuum as inside the bag and thus ensures
bound states.

With appropriately chosen parameters $a,b,c,B$ and $g_0$ Goldflam and
Wilets \cite{GoWi82} found in the mean field approximation (MFA)
localized static solutions in good agreement with
the experimentally known properties of the nucleon.
At this point, however, one already encounters a shortcoming of the model.
For any reasonable choice of the parameters the asymptotic mass of the quarks
is rather small, barely exceeding the mass inside the bag. Therefore there is
no real confinement in this model.
The solution of this problem was first given by
by Bayer et al. \cite{BaFoWe86} and F\'{a}i et al. \cite{FaPeWi88},
who replaced the original coupling term (\ref{eq:coupling})
by a nonlinear coupling function
\be
g_{eff}(\sigma) = g_0 \, \sigma_{vac}\left[\frac{1}{\kappa(\sigma)} - 1\right]
\ee 
using
\be
\kappa(\sigma) = 1 + \theta(x) x^n [nx -(n + 1)]  \label{theta}
\ee
\be
\mbox{with}\quad\quad x = \frac{\sigma}{\sigma_{vac}}\quad\quad \mbox{and} \quad
n = 1,2 {~\rm or~} 3
\ee
where $\kappa$ plays the role of the vacuum permeability in analogy with
the dielectric function in a polarizable medium. We use $n =1$.
$\kappa (\sigma)$ was chosen in such a way that
$
\kappa(\sigma) \rightarrow  0  \mbox{~for~} \sigma \rightarrow \sigma_{vac}
\mbox{~and~}
\kappa(\sigma) \rightarrow  1 \mbox{~for~} \sigma \rightarrow 0 .$
As a consequence $g_{eff} (\sigma)$ interpolates between the perturbative
and the nonperturbative vacuum sector.
The divergence of $\geff $ at $\sigma = \sigma_{vac}$ leads
to an infinite asymptotic mass of the quarks providing confinement this way.

In two subsequent papers
\cite{KrTaWiWi88},\cite{KrTaWiWi91} it was shown that this form of the effective coupling
gives a good approximation to the effects of the gluon field  in the
chromo-dielectric soliton model. Note
that this model, contrary to the Friedberg-Lee Lagrangian,
has no direct quark-$\sigma$ coupling term, and therefore
preserves chiral symmetry.
As shown in Ref. \cite{KrTaWiWi91}
the effective quark-$\sigma$ coupling arises after
dynamical chiral symmetry breaking in the so called local approximation.

By an appropriate choice of the parameters one is also in the extended
Friedberg-Lee model able to obtain static solutions which reproduce
the experimentally known properties of the nucleon as well as in the
standard model.
Based on this experience we formulate a transport theory with the
Friedberg-Lee Lagrangian
and investigate	both the original and the confining version in our simulations.

\section{Transport Theory} \label{transport}
In this section we present the transport
equation, which describes the motion of the quarks.
We follow here the spirit of
the well established BUU model \cite{Blattel},\cite{EVG}, which has
been sucessfully applied to the heavy-ion collisions.
A derivation of the relativistic transport equation has been extensivly discussed in
 the literature  \cite{Blattel,EVG,StHe93,BM90,LWK89} and the application to the
 Friedberg-Lee model was already given in \cite{KaVeBiMo93}.
Therefore  at this point we just sketch the derivation of the transport equation.
We introduce the Wigner-matrix
\be
\label{eq:defwig}
W_{\alpha \beta} (X,p) \stackrel{\rm def}{=}
\frac{-i}{(2 \pi)^4} \int d^4u \, e^{-ipu} \left( \bar{\Psi}_\beta (X+\frac{u}{2})
\Psi_\alpha (X-\frac{u}{2}) \right)_{\alpha \beta} \, .
\ee
This is the quantummechanical analogon of the classical phase space
distribution function.
After applying the Dirac-equation, a semiclassical expansion,a spinor decomposition of W and some lengthy algebra, one
arrives at
\be
W = (p_\mu \gamma^\mu + m^*) F(x,p)   \, ,
\ee
and
\bea
\label{eq:vl5}
(p_\mu \partial^\mu + m^* \partial_\mu m^* \partial^\mu_p ) F(x,p) &=& 0 \\
\label{eq:vl6}
(p_\mu p^\mu -(m^*)^2 ) F(x,p) &=& 0	  \, ,
\eea
with
\be
m^* = m_0 + \geff \, .
\ee
Equation (\ref{eq:vl6}), the so called mass-shell constraint, enables us
to reduce the eight-dimensional distribution function $F(x,p)$ to a
seven-dimensional hypersurface in phase space. If one
defines the positive energy solutions by
\be
F(x,p) = \frac{1}{p_0} \delta (p_0 - \omega) f({\bf{x}},{\bf{p}} ,t),
\ee
with
\be
\label{eq:defw}
\omega = \sqrt{p^2+(m^*)^2} \, ,
\ee
then equation (\ref{eq:vl5}) gives the well known Vlasov equation
\be
\label{eq:vlasov}
\left(\omega \partial_t + {\bf{p}} \nabla_x - m^* (\nabla_x m^*) \nabla_p
\right)
f({\bf{x}},{\bf{p}},t) = 0 \, .
\ee
This equation together with the Klein-Gordon equation
\be
\label{eq:kge}
\partial^2_t \sigma = \nabla^2_x \sigma -\frac{dU (\sigma)}{d \sigma}
 - \frac{d\geff} {d \sigma} \rho_S \, .
\ee
determines the time-evolution of a quark system in the present model.

For the numerical treatment of this equation we use the so called test particle ansatz for the
scalar phase-space distribution function $f({\bf x},{\bf p})$ \cite{Wang}:
\be
f(\vx,\vp) =
\sum_n \delta (\vx - \vx_n(t) ) \delta(\vp -
\vp_n(t) ) \quad . \label{eq:testpar1}
\ee
Inserting it into the Vlasov equation (\ref{eq:vlasov}) shows that the so
called test particles have to follow
classical trajectories according to Hamilton's equations of motion:
\bea
\dot{\vx_n} &=& \frac{\vp_n}{\omega_n} \label{eq:test1} \\
\dot{\vp_n} &=& - \frac{m^{\ast}_n}{\omega_n} \nabla_x m^{\ast}_n	\quad . \label{eq:test2}
\eea
with
\be
\omega_n = \sqrt{{\vp_n}^2 + {m^\ast}^2(\vx_n)}
\ee

The scalar and baryon density, expressed through the phase-space distribution $f({\bf{x}},
{\bf{p}}, t)$, are given as
\bea
\rho_S (x) &=& \int d^4p \, \,tr (W(x,p)) = \frac{\eta}{(2 \pi)^3} \int
\frac{d^3p}{\omega} m^* f({\bf{x}},{\bf{p}},t) \label{eq:rhos1}\\
\rho_B (x) &=& \int d^4p \, \, tr (\gamma_0 W(x,p)) = \frac{\eta}{(2 \pi)^3} \int
f({\bf{x}},{\bf{p}},t) \, .	\label{eq:rhob1}
\eea
Here we have introduced a degeneracy factor $\eta$, which accounts
for the internal degrees of freedom.
In our case  we assume
\be
\eta = 2_{\mbox{Spin}} \times 2_{\mbox{Isospin}}
\times 3_{\mbox{Color}} = 12.
\ee
In terms of the test particles these densities are expressed as
\bea
\rho_B (\vx) &=& \frac{Q}{N}\sum_{n=1}^{N} \delta (\vx -\vx_n) \\
\label{eq:rhostest}
\rho_S (\vx) &=& \frac{Q}{N}\sum_{n=1}^{N} \frac{m^*}{\omega_n} \delta (\vx -\vx_n)
\eea
where $N$ denotes the total number of test particles and $Q$ is the number of
valence quarks, in our case $Q=3$ for each nucleon.
%
%

In Reference \cite{KaVeBiMo93} we showed how antiparticles can be incorporated
in a transport equation in one dimension. This is, of course, in complete analogy
also possible in three dimensions and will be presented in a future work, in which
we are going to study pair-production.
At the present stage we assume
that the quark phase-space distribution function
represents the valence-quark distribution and that all sea effects
can be absorbed into the parameters of the phenomenological potential
$U (\sigma)$, as done in the quantum mechanical calculations \cite{GoWi82}.

\section{Static solution} \label{static}
\subsection{The radial soliton}
As  a prerequisite for a dynamical calculation we are faced with the
task to find the static solution of the transport equation
for the quark distribution function coupled to the static Klein-Gordon equation for
the $\sigma$-field.
In the static case the Vlasov equation (\ref{eq:vlasov}) and the Klein-Gordon
(\ref{eq:kge}) reduce to
\bea
{\bf{p}} \nabla_x  f({\bf{x}},{\bf{p}},t) &= &	m^* \nabla_x m^* \nabla_p
f({\bf{x}},{\bf{p}},t)
\label{eq:stat1}\\
\nabla_x^2 \sigma &= & \frac{d U(\sigma)}{d\sigma} + \frac{d \geff}{d \sigma}
\rho_s (m^*(\sigma))
\label{eq:stat2}
\eea
By replacing $\nabla_x \rightarrow (\nabla_x \, \omega) \, d/d \omega$
and $\nabla_p \rightarrow (\nabla_p \, \omega) \, d/d \omega$
it is easy to see that any arbitrary function $f (\omega)$ solves
equation (\ref{eq:stat1}), if $\omega$ is defined by
equation (\ref{eq:defw}). In addition  the
idempotency property of the single particle density matrix
$\rho^2 = \rho$ yields in the semiclassical limit $f^2 = f$ (see \cite{Lee90}),
restricting $f$ to be either 0 or 1. Both conditions together
yield the local Thomas-Fermi approximation for the distribution function
\be
f(\omega) = \theta (\mu - \omega) \, ,
\ee
where we have introduced the Fermi energy $\mu$.

This form enables us to express the scalar- (\ref{eq:rhos1}) as well as the
baryondensity (\ref{eq:rhob1}) as a function of $m^*$ only.
\bea
\rho_S (m^*) & = & \frac{1}{(2 \pi)^3} 2 \pi \eta m^* \left(
\mu {\bf{p_f}} + (m^*)^2 \ln(\frac{m^*}{\mu + {\bf{p_f}}}) \right) \theta (\mu
- m^*) \label{eq:rhos2} \\
\rho_B (m^*) & = & \frac{1}{(2 \pi)^3} \frac{4 \pi \eta}{3}
{\bf{p_f}}^3 \theta (\mu- m^*) \label{eq:rhob2} \,
\eea
with the fermi momentum  $ {\bf{p_f}} =\sqrt{\mu^2 - (m^*)^2} $.

Because the scalar density depends only implicitly, i.\ e.\ through $m^*$,
on $x$, the solution	of the coupled system	(\ref{eq:stat1}) and
(\ref{eq:stat2}) is reduced to the solution
of equation (\ref{eq:stat2}), in which the scalar density is given by (\ref{eq:rhos2}).
By introducing an effective potential satisfying
\be
\label{eq:ueff1}
\frac{ d U_{eff}}{d \sigma} = - \frac{d U}{d \sigma} - \frac{d \geff}{d \sigma} \rho_S
\ee
the stationary Klein-Gordon equation (\ref{eq:stat2}) can be cast in a form which
resembles a Newtonian equation of motion for
a fictitious particle with unit mass, coordinate "$\sigma$" and time "$r$",
moving in a Potential $U_{eff}$:
\be
\label{eq:stat3}
\nabla^2 \sigma = - \frac{ d U_{eff}}{d \sigma}
\ee
In \cite{Col85} it was shown
how this analogy can be used in one dimension to obtain the solution
of a field equation similar to (\ref{eq:stat2}).
If one assumes a spherically symmetric system, the classical analogy also
holds in three dimensions, the Laplacian yielding a time dependent friction term
in the analogous Newtonian equation
\be
\frac{d^2 \sigma}{d r^2}  = - \frac{2}{r} \frac{d\sigma}{d r} -
\frac{ d U_{eff}}{d \sigma} \, .
\ee
By integration of equation (\ref{eq:ueff1}) the effective potential is
given by
\be
\label{eq:ueff2}
U_{eff} = - U (\sigma) - \int \frac{d \geff}{d \sigma} \rho_S \, d \sigma
\ee
By substituting the scalar density (\ref{eq:rhos2})  into equation
(\ref{eq:ueff2}) and changing the integration variable from $\sigma$ to
$m^*$ the integral can be done analytically and the effective
potential is obtained as
\be
\label{eq:ueff3}
U_{eff} = - U (\sigma) + \frac{\eta \pi }{3 (2 \pi)^3}	\theta (\mu -
m^*) \left( \mu {\bf{p_f}}^3 - \frac{3}{2} \mu (m^*)^2 {\bf{p_f}} -
 \frac{3}{2} (m^*)^4 \ln (\frac{m^*}{\mu + {\bf{p_f}}}) \right) \, .
\ee
The solution of the coupled Vlasov equation and the stationary Klein-Gordon
equation has therefore been reduced to the equation of motion for a
classical particle moving in an effective potential $U_{eff}$.
The parameters of the potential $U (\sigma)$ as well as
the coupling constant and the Fermi energy are adjusted to reproduce the
experimentally
known baryon-number, mean mass of delta and nucleon, and rms radius of the nucleon.
They are summarized in table (\ref{tab1}), where 'L' denotes the original
model  and 'C' the confining version of the Friedberg-Lee model.
\begin{table} [ht]
\center{\begin{tabular}{||c|c|c||} \hline
parameter set & L & C	  \\ \hline
coupling	  & original	    & confining \\	\hline
$a$ $[fm^{-2}]$   & 0.0 	    & 20.52	 \\
$b$ $[fm^{-1}]$   & -420.0	    &-420.64	  \\
$c$ $[1]$	  & 4980.	    &2594.0	  \\
$B$ $[fm^{-4}]$   & 0.283	    &0.1269	  \\
$g$ $[1]$	  & 8.0 	    &1.139	   \\
$m_0$ $[fm^{-1}]$ & 0.025	    &0.025	 \\
$\mu$ $[fm^{-1}]$ & 1.768	    &1.70238	  \\
$\eta$		   & 12 	    &12 	\\
$\sigma_{vac}$ $[fm^{-1}]$ & 0.253  &0.3514	\\
\hline
\end{tabular}}
\caption{The table shows the parameters used in the three-dimensional model.}
\label{tab1}
\end{table}
The effective potential in case of the linear coupling is displayed in
the upper panel of figure 1
and for the confining model in the lower panel of
figure 1.
In order to obtain a soliton solution	the
$\sigma$ field should arrive at its vacuum value for $r \rightarrow \infty$.
In addition for a regular solution we also require $d \sigma / d r|_{r=0} = 0$.
This means the
$\sigma$-particle starts with a "velocity" $d \sigma/dr = 0$ at time
"$r=0$", reaches a maximum velocity and finally comes to rest at $\sigma =
\sigma_{vac}$. In the three-dimensional case,
due to the friction term the $\sigma$-particle has to start  a little bit
above the barrier, having thus more potential energy.
If the starting height is too large the particle escapes
to infinity, if it is too low one obtains periodic solutions for the
soliton field.

Solving the equations of motion for the $\sigma$-particle and choosing
the initial value $\sigma (r=0)$  such that $\sigma (r=\infty) = \sigma_{vac}$
we are able to find solitonic solutions.
The $\sigma$-field corresponding to this motion together
with the baryon- and scalardensity is displayed in figure 2 in case of
the original Friedberg-Lee model,
where we have scaled the $\sigma$-field with a factor of 5 and multiplied the
scalar density with $g_0$.

We find a volume centered baryon density, whereas the scalar density is surface
peaked. A similar behavior is seen in figure 3,
in which the same quantities are shown for
the confining model. However note the rapid fall-off of the densities
at the surface, which is due to the strong rise of the effective mass in this region.
Table (\ref{tab2}) shows the results of the fits compared to the
experimental data.

\begin{table} [ht]
\center{\begin{tabular}{||c|c|c|c||} \hline
Parameter set & L & C & Experimental Data \\ \hline
$B$ $[1]$	   & 1.0	    &1.0	&  1.0	\\
$E$ $[MeV]$	   & 1101	    &1140	&  1087 \\
$RMS$ $[fm]$	   & 0.67	    &0.674	&  0.83 \\
$Gluemass$ $[MeV]$	   & 1434	    &1129	& $\approx 1500	$\\ \hline
\end{tabular}}
\caption{The table shows the results
of the fits for parameterset L and C
compared to the experimental  data. The experimental data
are taken from \protect\cite{Wilbuch}.}
\label{tab2}
\end{table}

We find a good agreement with the observed properties of the nucleon,
although the radius comes out somewhat too small. This is a common feature of
the semiclassical approximation, because here the density can not exceed
the classically allowed region, whereas in the quantum mechanical treatment the
long tail of the wavefunction accounts for a larger radius.

It is also worth mentioning that an effective potential, as we have
constructed above, allows the determination  of those regions of the
parameter space,
where soliton solution can exist. This is, because a sufficient height
of the barrier at $\sigma = 0$ is needed to overcome the friction-term.
Shenghua and Jiarong \cite{ShJi93} presented a model in which the potential parameters in $U (\sigma)$
are temperature dependent in order to model a soliton in a background of
a quark gluon plasma. In that case the effective potential allows
for the determination of a phase-transition.

\subsection{Stability of the static solution}
In the last section we have reduced the solution of the coupled Vlasov equation
and the Klein-Gordon equation to the solution of equation (\ref{eq:stat3}),
which contained only the scalar field. From Derrick's theorem \cite{De64}, however,
one would naively expect that no stable soliton should exist in three dimensions.
This conclusion is reached by studying the energy under
scaling transformations $\hat{r} = \lambda r$. For a static solution
it is a necessary condition that the first variation of the energy with
respect to
the scaling parameter $\lambda$ vanishes and the second variation is larger than 0. Derrick's
theorem states that for a scalar theory this conditions can't be
simultaneously met in three dimensions. A first way out from this dilemma
is to add a fourth-order
gradient term to the Lagrangian, as it is done in the Skyrme
model \cite{Sk62,ZaBr86}. Also a second possibility to circumvent Derrick's
theorem exists however, if there is an additional conserved quantity (see for example
\cite{LePa92}).
This is the case in our model; the Vlasov equation, as it easily seen,
conserves the number of quarks
\be
Q = \int \int d^3x d^3p \, \, f({\bf x},{\bf p}) \, .
\ee
This condition implies that under a naive scale transformation,
$\hat{r} = \lambda r$,
the quark number
changes to
\be
Q (\lambda) =\frac{Q(1)}{\lambda^3}    \,
\ee
violating the baryon number conservation law.
Therefore we have to consider restricted scaling transformations, which
preserve the baryon number.  Respecting this constraint we adjust the
Fermi energy $\mu$ to $\mu (\lambda)$, so that for all values of the scaling variable
$Q (\lambda) = Q (1) = 3.$ In figure 4
the energy of the soliton is shown as a function of the scaling variable
$\lambda$ under the constraint of baryon number conservation for the
confining model.
This figure clearly shows an energy increase if the system is compressed or
expanded, proving thus the stability of the soliton solution.
The same argument also holds for the original model.

We also have checked the stability of the system dynamically, solving the
time-dependent Vlasov  and  Klein-Gordon equations by methods
described in the Appendix. We tested the response of the system if the
soliton is initially compressed by a factor of $0.9$\,. As shown in figure 5
we observe an oscillation of the soliton radius. The damping of
the amplitude can be attributed to a numerical viscosity effect (see appendix).
The oscillation frequency of this monopole mode corresponds to an excitation
energy of 377 MeV, which could be associated with the 1440 MeV Roper
resonance.  As a further test of the stability
we have initialized the static solution of equation (\ref{eq:stat3}) and
evolved the system numerically.
We checked for time scales up to 18 fm/c, that the density distribution doesn't
change.
This is clearly sufficient to simulate
collisions for a time of the order of 3-4 $fm/c$.
We conclude that in the original and in the confining Friedberg-Lee model there
are semiclassical static solutions in
reasonable agreement with the experimentally known properties of the
nucleon.
This solution is also stable, therefore can be used as	an initialization
in a dynamical treatment of nucleon-nucleon collisions.

\section{Collisions} \label{collisions}

In order to simulate collisions	between two nucleons we proceed in the following way.
First we initialize two nucleons according to the method described in section
\ref{static}, then we boost them by the procedure already described in \cite{KaVeBiMo93}.
As the observer's frame we choose the center of mass of the two nucleons.
The simulations are all done using the test particle method for the Vlasov equation.
The resulting Hamiltonian equation of motion (\ref{eq:test1}) and
(\ref{eq:test2}) as well as the mean field equation (\ref{eq:kge})  are
integrated by a staggered leapfrog method. Details of the numerical
implementation are given in the Appendix.

We have calculated collisions
in the energy range of 0.1 - 20 GeV bombarding energy in the
laboratory frame and studied central as well as very peripheral collisions.
Initially the nucleons are well separated in coordinate space and we follow
the time evolution until they are separated again. For energies above
0.2 GeV we observe that the nucleons penetrate each other, overlap
for a certain time, in which the baryon density almost doubles, and then
separate again and leave the collision zone. A short time after separation the
original shape of the solitons is restained.

For a central collision of two
nucleons with a bombarding energy  of
3.98 GeV the time evolution of the baryon density in the reaction
plane is shown in figures 6
and 7
in case of the confining model.
One sees clearly the penetration
of the two nucleons; notice, however, the different scales of the
z-axis in the figures.	It is also seen	that in the
overlap zone almost twice the initial density is reached. It is also noticeable that
the nucleons are still somewhat	distorted when they leave the
reaction zone at $t=3.2 fm/c$; however, at $t=8 fm/c$ they regain
their original shape. Therefore it is necessary to follow the time evolution
for some time, even after the nucleons are separated.

The time evolution of the confining mean field
potential can be seen in figures 8
and 9
where the scalar field
as well as the scalar and baryon density along the beam axis are displayed for
the same central collision.
In the moment the two nucleons make contact the density pushes
the $\sigma$-field down, thus creating a 6-quark bag at $t=0.8	fm/c$.
Notice again that the quark-density reaches $2.5/fm^3$ compared to the
initial $1.6/fm^3$ of a single nucleon. Then the $\sigma$-field overshoots
to small negative values at $t =2.4 fm/c$.

Due to
the large momentum of the soliton field the two bags penetrate each other so
fast that the confining wall between the bags is reestablished before
the first quarks (test-particles) reflected from the outer wall reach the
central region between the two bags. This is seen at timestep $t=3.2 fm/c$.
This reestablishing of the confining wall between the nucleons leads to
the confinement of the quarks inside the two separated bags. The
$\sigma$-field in the intermediate region swings then back to the vacuum value.
As already mentioned this process needs some time so that only after
$t= 5-6 fm/c$ the $\sigma$-field approaches its asymptotic shape. The baryon
density still deviates a little from the initial configuration, however
 after another $2 fm/c$ the original shape is also resumed.

Figure 10
and 11
show the time evolution
in phase space ( $p_z$ versus $z$ ) for a central collision at a bombarding energy of 2.88 GeV in
the laboratory frame.	It is
recognizable that at that energy the nucleons still overlap considerably
with their momentum distributions as can be seen from the picture for $t =0 fm/c$.
As in the one-dimensional
case  this effect is due to the small quark rest mass in
the middle of the bag. Shortly after the two nucleons overlap the phase space
distribution of the nucleons gets distorted, but there is no substantial
interaction in phase space. After about $8 fm/c$ the original phase space
distribution of the nucleons is restored.
Notice, however, that the center of the
phase-space distribution of target and projectile have slightly approached
each other in momentum space.
The energy as  a function of time is shown in figure 12,
where a central collision with a bombarding energy of 2.88 GeV is
considered. Besides the total energy the figure also shows the energy,
attributed to the quarks
\begin{equation}
E_q = \int \, \int \, \omega f(\vx,\vp) d^3x \, d^3p
\end{equation}
and the contribution of the $\sigma$-field.
\begin{equation}
E_\sigma = \int \, \int \, \left( \frac{1}{2} (\partial_t \sigma)^2
+ \frac{1}{2} (\nabla \sigma)^2 + U(\sigma) \right) d^3x \, d^3p
\end{equation}
The total energy loss after $t = 10 fm/c$ amounts to 7 \%, which is due
to a numerical viscosity term, implemented in the numerical scheme as
explained in the appendix.
Notice, however, that during the time  the two nucleons overlap, there
is a considerable exchange of energy stored in the mean-field and in the
fermionic part. After the collision only some small oscillations in the
different energy contributions remain.

In case of the original model we see that the time evolution of a collision
is very similar to the confining model.	As an example figure 13
shows the baryon density for a collision with an
impact parameter of 1.6 fm in the reaction plane. One recognizes how
the nucleons form a common bag, turn slightly around each other and then
leave the reaction zone essentially unchanged.

For very low bombarding energies up to 0.2 GeV we observe a fusion of the bags. The
same behavior is already seen in the one-dimensional calculations, although
there we observed fusion up to beam energies of 1 GeV. This fusion can be
explained as follows.
For small energies the soliton bags move much more slowly than the
quarks, which, in the middle of the bags, move almost with the speed of light.
Exactly as in the previously considered case of higher energies the density pushes the
$\sigma$-field between the solitons down. However low bombarding energies
result in a low rated change of the mean-field, and the first quarks reflected from
the outer soliton wall reach the intermediate region between the bags before
the $\sigma$-field had time to be build up again. The quarks then spread
over the entire 6 quark bag and a bag fusion occurs. This is shown in
figure 14
by a time
evolution plot of the baryon density in the reaction plane. The collision considered here is a central
collision with a bombarding energy of 90 MeV in the laboratory frame.
The solitons first merge and afterwards the fused bag oscillates
 at times $t =2.4 - 14.4 fm/c$.
We note, that Schuh et. al. \cite{Schuh}
estimated the potential
from nucleon nucleon scattering in the Friedberg-Lee model in a
Born-Oppenheimer (adiabatic) approximation and also end up with an attractive
potential, which is responsible for the fusion of slowly moving bags.
This nucleon fusion is an artifact due to the neglect of repulsive
forces	in the present model.

In the following we study the trajectories of the nucleons. In order
to define target and projectile throughout the entire collision we employ
the following procedure. For the initial state of two well separated nucleons
we construct a plane located in the middle of the vector connecting the
centers of mass of target and projectile. The orientation of the plane is
perpendicular to this connecting line. Then, in a next step, the center of mass
of the baryon distribution on both sides
of this plane is calculated seperately. The new center of mass coordinates
are then used in the next timestep to construct the separating plane again. In this
way we are able to define target and projectile throughout the entire collision.

Figure 15
shows the	center of momentum
trajectory of the target in the reaction
plane for four different impact parameters. The bombarding energy in the
laboratory frame was in this case 3.98 GeV and the confining model was used.
We see for all but zero impact parameters a rather
strong deflection of the nucleon, which starts with a z-coordinate $z= 1.25 fm$
and then moves to the left.
Since $z=0 fm$ corresponds  to a region
where the nucleons overlap and form for a while a common bag, at this
point target and projectile can't be defined in a proper way and the
definition of projectile and target described above leads to a bump
in the projectile coordinate.
Notice also the deflection towards negative $x$-values implying a negative deflection
angle, as one would expect from an attractive potential. The figure also shows
that the maximum deflection angle is reached for intermediate impact parameters.

This behavior is also seen in figure 16,
where the deflection
angle is shown as a function of the impact parameter for five different
bombarding energies. The figure shows the results of a simulation done in
the confining model. With increasing beam energy the deflection angle
decreases, therefore at very high energies ($> 10$ GeV) the nucleons become almost
transparent.  This is a direct consequence of the decreasing
interaction time  with increasing beam energy.
There is also a shift in the impact parameter belonging to the maximum
deflection angle to
smaller values with increasing beam energy visible.
At low energies the largest deflection angle is found for an impact-parameter
of $1.2 fm $, implying a greasing reaction.
In addition figure 16 shows the deflection angle for a bombarding energy
of 3.98 GeV in the original nonconfining model. One notes a similar behavior
as in the confining model, however, the deflection is
less than in the confining model and ocurs at a lower impact-parameter.

If one considers the quarks exchanged between projectile and target,
a very similar dependence on the beam energy as in case of the deflection angle
is found. This is displayed in figure 17,
which shows
the exchanged baryon density as a function of impact parameter for six
different bombarding energies in the confining model. For low bombarding energies
we find a substantial quark exchange up to 10 \% for a collision with an
intermediate impact parameter. As already mentioned above in this scenario
the baryons almost surround each other and therefore stick together
for a rather long period of time, which in turn explains  the large
quark exchange. At higher bombarding energies, however, the nucleons become
almost transparent and the exchanged baryon density negligible.
The original model shows	for the same bombarding energy less exchanged
quark density consistent with the behavior of the deflection angle
shown above.

\section{Summary and Conclusion} \label{summary}

In this paper we have described a semiclassical transport theory on the quark
level which includes some
nonperturbative aspects of QCD. This is done by including a soliton mean--field which 
governs the dynamics. As a starting point we chose a phenomenological quark model for 
the nucleon, namely the Friedberg--Lee model. We have derived equations of motion for the 
time evolution of the Wigner--function which leads in the semiclassical expansion to the 
well known Vlasov equation combined with a mass--shell constraint. For the Wigner--
function we have used a test particle ansatz which results in classical equations of motion.

Considering different extensions to the original Friedberg--Lee model it turned out that
the formulation with an effective quark--$\sigma$ coupling, which comprises
many--gluon--exchange effects, can be used to model confinement on the test particle
level. Inside the dynamically generated bag the quarks are nearly massless (`asymptotic 
freedom'). 

For the initialization of a stable nucleon we have used a consistent solution of the stationary 
Vlasov equation. It was shown that a local Thomas-Fermi distribution which depends only on
energy and not explicitly on position and momentum is the correct semiclassical
solution of the stationary Vlasov equation. After the
construction of the semiclassical
nucleon by means of an effective potential we have studied the stability of
the static solutions in the original and in the confining model.
We have shown how the quark number conservation due to the Vlasov equation
guarantees the stability of the static solutions.

Having all these ingredients we have performed calculations of collisions in 3+1 dimensions.
We have studied various impact parameters and
bombarding energies.
It has turned out that on
the pure mean field level the nucleonic bags are transparent at high bombarding energies, and
the confinement is realized in each nucleon separately. For very low bombarding energies
($ 90 MeV$ in the lab.\ frame) we observe a fusion of the bags with a following oscillation
of the six quark bag. This is an artifact of the approximation used.

Summarizing, these simulations have shown that the semi\-clas\-si\-cal
Fried\-berg-Lee model is able to des\-cribe moving quark
systems and confinement in a collision of nucleons. This model is
an excellent
basis for extensions including direct quark--quark collisions, on the
top of the stable colliding nucleons we describe so far.

\newpage
\section*{Appendix}
\label{app}
In this appendix we explain some details of the numerical implementation
of our simulation. As already mentioned we use for the time integration of
the Vlasov equation the test particle method of Wong \cite{Wang}, which
results in solving the Hamiltonian equation of motion (\ref{eq:test1})
and (\ref{eq:test2}) for the test particles. As it is usally done in these
simulations we define the $\sigma$-field, $\dot{\sigma}$, and the densities
on a spatial
grid, wheras the test particle coordinates are continous variables.
In our actual simulation we use a 41 x 41 x 129 grid with a
typical grid spacing of 0.1 - 0.15 fm in transverse direction and
${0.1 {\rm fm}}/\gamma$	in the boost direction. The number of test particles used
was between 100.000-200.000.
The time-integration of the $\sigma$-field as well as of the test particles
is done by a staggered leapfrog method as described in
\cite{Treesph}.
\bea
\label{eq:up1}
{\vx_n}^{t+2 \triangle t} & = & \vx_n^t + 2 \, \triangle t \,
{\frac{\vp_n}{\omega_n}}
^{t+\triangle t} + O(\triangle t^3) \\
{\sigma}^{t+2 \triangle t} & = & \sigma^t + 2 \, \triangle t \, {\dot{\sigma}}^
{t + \triangle t} + O(\triangle t^3)\\
\vp_n^{t+3 \triangle t} & = & \vp_n^{t + \triangle t} + 2 \, \triangle t \,
{\frac {-m^* \nabla_x m^*}{\omega_n}}^{t+ 2 \triangle t} + O(\triangle t^3)\\
{\dot{\sigma}}^{t+3 \triangle t} & = & {\dot{\sigma}}^{t +  \triangle t}
+ 2 \, \triangle t \,	{\rm rhs}^{t+ 2 \triangle t} + O(\triangle t^3) \, ,
\eea
where the superscripts refer to the time step at which the quantities
are evaluated and "rhs" denotes the right hand side of equation (\ref{eq:kge}).
However, the test particle velocity depends through $\omega$ on the
$\sigma$-field and on the test particle coordinate. In this case
second order accuracy is only preserved if we assign first a tentative
coordinate and $\sigma$-field as
\bea
\vx_n^{t+ \triangle t} & = & \vx_n^t + \triangle t \,
{\frac{\vp_n}{\omega_n}}
^{t- \triangle t} \\
\sigma^{t+ \triangle t} & = & \sigma^t + \triangle t \, {\dot{\sigma}}^{t -
\triangle t}  \\
\vp_n^{t+2 \triangle t} & = & \vp_n^{t + \triangle t} +  \triangle t \,
{\frac {-m^* \nabla_x m^*}{\omega_n}}^{t} \\
{\dot{\sigma}}^{t+2 \triangle t} & = & {\dot{\sigma}}^{t +  \triangle t}
+  \triangle t \, {\rm rhs}^{t} \, .
\eea
and then in turn allow the coordinate to be updated according to
equation (\ref{eq:up1}).
This consistent update of $\sigma$-field and test particle coordinates and
momenta is absolutely necessary for a proper propagation, because the
test particle velocity is especially in the surface region very sensitive
to the exact test particle coordinate.	The timestep width was chosen
to be $1/10$ of the spatial gridspacing in order to fullfill the
Courant condition \cite{Numerical}.

To calculate the right hand side of equation (\ref{eq:kge}) the
scalar-density is calculated according to equation (\ref{eq:rhostest}).
However, due to the limited number of test particles  there are still
statistical fluctuations
in the density. The densities are therefore smoothed additionally by
letting the test particles contribute also to the neighboring cells
\cite{Lang}.
In addition to this we also smooth $\dot{\sigma}$ after every 25 timesteps,
in order to supress the high frequency fluctuations. This standard procedure
in fluid dynamics \cite{Treesph} causes the
numerical dissipation which in turn is responsible for the
energy loss seen in figure 12.

\newpage
\pagestyle{empty}
\section*{Figure captions}
\bigskip \noindent
{\bf Fig 1:~}
\label{fig:uefflin}
The effective potential $U_{eff}$ (solid line)
as a function of $\sigma/\sigma_{vac}$.
In addition contribution of $U(\sigma)$ (dashed line) and the
part of the effective potential stemming from the
scalar density (dotted line) are shown. The upper panel shows the potential
for the original Friedberg-Lee model, whereas in the lower panel the potential
in case of the confining model is displayed.

\bigskip \noindent
\label{fig:statlin}
{\bf Fig 2:~}
Baryon density (dashed), scalar density multiplied with $g_0$
(dotted) and $\sigma$--field (solid)
of the Friedberg--Lee model with linear coupling in three dimensions.
The parameter set L is used and the $\sigma$--field is
scaled by a factor of 5.

\bigskip \noindent
\label{fig:stateff}
{\bf Fig 3:~}
Baryon density (dashed), scalar density multiplied with $g'_{eff} (\sigma)$
(dotted) and $\sigma$--field (solid)
of the modified Friedberg--Lee model with nonlinear coupling in three dimensions.
The parameter set used is C and the $\sigma$--field is
scaled by a factor of 5.

\bigskip \noindent
\label{fig:lameff}
{\bf Fig 4:~}
The energy of the soliton as a function of the scaling parameter $\lambda$
in the confining model.

\bigskip \noindent
\label{fig:sh2rms}
{\bf Fig 5:~}
The root mean square radius  of the soliton, which is initalially squeezed
by a factor of 0.9, as a function of time.

\bigskip \noindent
\label{fig:ze80f0y3}
{\bf Fig 6:~}
Time evolution of the quark distribution of the confining model with an
impact parameter
$b = 0.0 fm$ and a bombarding energy of 3.98 GeV in the laboratory frame.
The quark-density is shown in the x-z (reaction) plane
from time $t = 0.0 fm/c$ up to $t =4.0 fm/c$.
Each conturline corresponds
to an increase of 0.183 $fm^{-3}$ in the density starting at 0.183 $fm^{-3}$.

\bigskip \noindent
\label{fig:ze80f0y4}
{\bf Fig 7:~}
Time evolution of the quark distribution of the confining model with an impact
parameter
$b = 0.0 fm$ and a bombarding energy of 3.98 GeV in the laboratory frame.
The quark-density is shown in the x-z (reaction) plane
from time $t = 4.8 fm/c$ up to $t = 13.6 fm/c$.
Each conturline corresponds
to an increase of 0.183 $fm^{-3}$ in the density starting at 0.183 $fm^{-3}$.

\bigskip \noindent
\label{fig:ae80f0y1}
{\bf Fig 8:~}
Time evolution of the quark distribution of a central collision
$b = 0.8 fm$ and a bombarding energy of 3.98 GeV in the laboratory frame.
The scalar (solid) and baryon density (dashed) and the $\sigma$-field (dotted)
along the beam axis are shown for times ranging from $t= 0.0 fm/c$ to
$t = 4.0 fm/c$. The parameter set used is C.

\bigskip \noindent
\label{fig:ae80f0y2}
{\bf Fig 9:~}
Time evolution of the quark distribution of a central collision
$b = 0.8 fm$ and a bombarding energy of 3.98 GeV in the laboratory frame.
The scalar (solid) and baryon density (dashed) and the $\sigma$-field (dotted)
along the beam axis are shown for times ranging from $t= 4.8 fm/c$ to
$t = 13.6 fm/c$. The parameter set used is C.

\bigskip \noindent
\label{fig:ze80f0p1}
{\bf Fig 10:~}
Time evolution of the phase-space distribution of the confining model
of a central collision
and a bombarding energy of 3.98 GeV in the laboratory frame.
The phase space density $p_z$ versus $z$ is shown
at times from $t = 0.0 fm/c $ to $t = 4.0 fm/c$.
Each conturline corresponds
to an increase of 0.23 starting at 0.23.

\bigskip \noindent
\label{fig:ze80f0p2}
{\bf Fig 11:~}
Time evolution of the phase-space distribution of the confining model
of a central collision
and a bombarding energy of 2.88 GeV in the laboratory frame.
The phase space density $p_z$ versus $z$ is shown
at times from $t = 4.8 fm/c $ to $t = 13.6 fm/c$.
Each conturline corresponds to an increase of 0.23 starting at 0.23.

\bigskip \noindent
\label{fig:en75i0}
{\bf Fig 12:~}
The total energy (solid line), the quark contribution (dotted line), and
the mean field contribution (dashed line) are shown as a function of time
belonging to a central collision with  $\protect{\sqrt{s}} = 3.4 GeV $.

\bigskip \noindent
\label{fig:zb75i8y}
{\bf Fig 13:~}
Time evolution of the quark distribution of the original model with an impact parameter
$b = 0.8 fm$ and a bombarding energy of 2.88 GeV in the laboratory frame.
The quark-density is shown in the x-z (reaction) plane
from time $t = 0.0 fm/c$ up to $t = 3.2 fm/c$.
Each conturline corresponds
to an increase of 0.3 $fm^{-3}$ in the density starting at 0.3 $fm^{-3}$.

\bigskip \noindent
\label{fig:ze30f0y1}
{\bf Fig 14:~}
Time evolution of the quark distribution of the confining model with an
impact parameter
$b = 0.0 fm$ and a bombarding energy of 0.09 GeV in the laboratory frame.
The quark-density is shown in the x-z (reaction) plane
from time $t = 0.0 fm/c$ up to $t =7.2 fm/c$.
Each conturline corresponds
to an increase of 0.188 $fm^{-3}$ in the density starting at 0.188 $fm^{-3}$.

\bigskip \noindent
\label{fig:te}
{\bf Fig 15:~}
The trajectory of the projectile in the reaction plane for a
collision with a bombarding energy of 3.98 GeV in the laboratory frame.
The nucleon starts from the right and moves to the left. The parameterset
used is C.

\bigskip \noindent
\label{fig:efftheta}
{\bf Fig 16:~}
The  deflection angle as a function of the impact parameter for
five different energies. All energies are bombarding energies in the
laboratory frame and all, but one, calculations were done using
the confining model. The crosses mark the deflection angle of a simulation
with a bombarding energy of 3.98 GeV done in the original
nonconfining Friedberg-Lee model.

\bigskip \noindent
\label{fig:effrbe}
{\bf Fig 17:~}
The  exchanged quark density as a function of the impact parameter for
five different energies. All energies are bombarding energies in the
laboratory frame. The confining model is used.


\newpage

\begin{thebibliography}{}

\bibitem{CaMo90} W. Cassing and U. Mosel,  Prog. Part. Nucl. Phys. {\bf 25} (1990) 1

\bibitem{CaMoMe} W. Cassing, V. Metag, U. Mosel, and K Niita,  Phys. Rep. {\bf 188} 
(1990) 363

\bibitem{SoStGr89/1} H. Sorge, H. Stcker and W. Greiner,  Ann. Phys. {\bf 192} (1989) 
266

\bibitem{SoStGr89/2} H. Stcker and W. Greiner,  Phys. Rep. {\bf 137} (1986) 277

\bibitem{Bertsch} G. Bertsch and S. Das Gupta,  Phys. Rep. {\bf 160} (1988) 189

\bibitem{Blattel} B. Blttel, V. Koch, and U. Mosel, Rep. Prog. Phys. {\bf56} (1993) 1

\bibitem{Boter} W. Botermans and R. Malfliet, Phys. Rep. {\bf198} (1990) 115

\bibitem{Elze1} H.--Th. Elze, M. Gyulassy and D. Vasak, Phys. Lett. B
{\bf 177} (1986) 402

\bibitem{Elze2} H.--Th Elze, and U. Heinz, in: Quark--Gluon Plasma--Advanced Series on
Directions
in High Energy Physics--Vol.6, ed. R. C. Hwa (World Scientific, Singapore 1990) 117

\bibitem{Stan} S. Mr\'{o}wczy\'{n}ski, in: Quark--Gluon Plasma--Advanced Series on
Directions
in High Energy Physics--Vol.6, ed. R. C. Hwa (World Scientific, Singapore 1990) 185

\bibitem{AnGuNi87} B. Andersson, G. Gustafson, B. Nielsson-Almquist, Nucl. Phys. {\bf 
B281} (1987) 289

\bibitem{We89} K. Werner,  Z. Phys. {\bf C42} (1989) 85

\bibitem{SoStGr89/2} H. Sorge, H. Stcker and W. Greiner,  Nucl. Phys. {\bf A498} 
(1989) 
567c

\bibitem{WaGy91} X.-N. Wang and M. Gyulassy,  Phys. Rev. {\bf D44} (1991) 3501

\bibitem{GeMu92} K. Geiger and B. Mller,  Nucl. Phys. {\bf B369} (1992) 600

\bibitem{KaVeBiMo93} U. Kalmbach, T. Vetter, T.\ S.\ Bir\'{o} and U. Mosel,
Nucl. Phys. {\bf A563} (1993) 584

\bibitem{Wil} W.--M. Zhang and L. Wilets,  Phys. Rev. {\bf C45} (1992) 1900

\bibitem{VeWaWaWy87} J. J. M. Verbarschoot, T. S. Walhout, J. Wambach and
H. W. Wyld, Nucl. Phys. {\bf A461} (1987) 603

\bibitem{AlKoSeSo87} A. E. Allder, S. E. Koonin, R. Seki, and
H. M. Sommermann, Phys. Rev. Lett. {\bf 59} (1987) 2836

\bibitem{KuPiZa93} A. Kudrayavtsev, B. Piette, and
W. J. Zakrzewski, Z. Phys. {\bf C60} (1993) 731

\bibitem{Wilbuch} L. Wilets Nontopological Solitons, Lecture Notes in Physics,
Vol. 24 (World Scientific, Singapore 1989)

\bibitem{FbLe77} R. Friedberg and T.D. Lee,  Phys. Rev. {\bf D15}(1977), 1694 und {\bf 
D16} 
(1977) 1096

\bibitem{FbLe78} R. Friedberg and T.D. Lee,  Phys. Rev. {\bf D18} (1978) 2623

\bibitem{GoWi82} R. Goldflam and L. Wilets, Phys. Rev. {\bf D25} (1982) 1951

\bibitem{WiBiLuHe85} L. Wilets, M.C. Birse, G. Lbeck and E.M. Henley, Nucl. Phys. 
{\bf 
A434} (1985) 129

\bibitem{LuBiHeWi86} G. Lbeck, M.C. Birse, E.M. Henley and L. Wilets,  Phys. Rev. 
{\bf D33} (1986) 234

\bibitem{LuHeWi87} G. Lbeck, E.M. Henley and L. Wilets, Phys. Rev. {\bf D35} (1987) 
2809

\bibitem{DeGoHeWi83} J.--L. Dethier, R. Goldflam, E.M. Henley and L. Wilets,  Phys. 
Rev. 
{\bf D27} (1983) 2191

\bibitem{Le79} T. D. Lee,  Phys. Rev. {\bf D19} (1979) 1802

\bibitem{BiBiMaWi85} M. Bickebller, M.C. Birse, H. Marschall and L. Wilets,
Phys. Rev. {\bf D31} (1985) 2892

\bibitem{BiBiWi87} M. Bickebller, M.C. Birse and L. Wilets,
 Z. Phys. {\bf A326} (1988) 89

\bibitem{BaFoWe86} L. Bayer, H. Forkel and W. Weise,  Z. Phys. {\bf A324} (1986) 365

\bibitem{FaPeWi88} G. F\'{a}i, J. Perry and L. Wilets,	Phys. Lett. {\bf B208} (1988) 1

\bibitem{KrTaWiWi88} G. Krein, P. Tang,  L. Wilets, and A. G. Williams
Phys. Lett. {\bf B212} (1988) 362

\bibitem{KrTaWiWi91} G. Krein, P. Tang,	L. Wilets, and A. G. Williams
Nucl. Phys. {\bf A523} (1991) 548

\bibitem{EVG} H.--T. Elze, D. Vasak, M. Gyulassy, H. Heinz, H. Stcker, W. Greiner,
Mod. Phys. Lett {\bf 2} (1987) 451

\bibitem{StHe93} S. Mr\'{o}wczy\'{n}ski and U. Heinz, Ann. Phys. {\bf 229} (1994) 1

\bibitem{LWK89} Q. Li, J. Q. Wu, and C. M. Ko, Phys. Rev. {\bf C39} (1989) 849

\bibitem{BM90} W. Botermans and R. Malfliet, Phys. Rep. {\bf 198} (1990) 115

\bibitem{Da84} P. Danielewicz, Ann. Phys. {\bf 152} (1984) 239

\bibitem{Wang} C. Y. Wong, Phys. Rev. {\bf C25} (1982) 1460

\bibitem{Lee90} S. J. Lee, Phys. Rev. {\bf C42} (1990) 610

\bibitem{Col85} S. Coleman, {\em Aspects of Symmetry}, Cambridge University
Press, Cambridge, 1985

\bibitem{ShJi93} D. Shenghua and L. Jiarong, Phys. Lett. {\bf B302} (1993) 279

\bibitem{De64} G. H. Derrick, J. Math. Phys. {\bf 5} (1964) 1252

\bibitem{Sk62} T. H. R. Skyrme, Nucl. Phys.{\bf A31} (1962) 556

\bibitem{ZaBr86} I. Zahed and G. E. Brown, Phys. Rep. {\bf 142} (1986) 1

\bibitem{LePa92} T. D. Lee and Y. Pang, Phys. Rep. {\bf 221} (1992) 251

\bibitem{Schuh} A. Schuh, H.J. Pirner, and L. Wilets, Phys. Lett. B {\bf 174} (1986) 10


\bibitem{Treesph} L. Hernquist, and N. Katz, Astrophysical Journal Supplement
Series {\bf 70} (1989) 419

\bibitem{Numerical} W. H. Press, S. A. Teukolsky, W. T. Vetterling, and
B. P. Flannery, {\em Numerical Receipes}, Second Edition,
Cambridge University Press, Cambridge, 1992

\bibitem{Lang} A. Lang, PhD thesis, Univ. of Giessen, 1991, unpublished
\end{thebibliography}
\end{document}